\documentclass{Interspeech2024}
\usepackage{cite}
\usepackage{hyperref}
\usepackage{tablefootnote}
\hypersetup{
    colorlinks=true,
    linkcolor=red,
    urlcolor=magenta,
    }

\usepackage{lipsum}
\usepackage{soul}
    
\interspeechcameraready

\title{Can Large Language Models Understand Spatial Audio?}

\name[affiliation={1,*}]{Changli}{Tang}
\name[affiliation={1,*}]{Wenyi}{Yu}
\name[affiliation={2}]{Guangzhi}{Sun}
\name[affiliation={3}]{Xianzhao}{Chen}
\name[affiliation={3}]{Tian}{Tan}
\name[affiliation={3}]{\\Wei}{Li}
\name[affiliation={3}]{Jun}{Zhang}
\name[affiliation={3}]{Lu}{Lu}
\name[affiliation={3}]{Zejun}{Ma}
\name[affiliation={3}]{Yuxuan}{Wang}
\name[affiliation={1,\clubsuit}]{Chao}{Zhang}

\address{
  $^1$Department of Electronic Engineering, Tsinghua University\\
  $^2$Cambridge University Engineering Department \\
  $^3$ByteDance}
\email{\{tcl20,ywy22\}@mails.tsinghua.edu.cn, cz277@tsinghua.edu.cn}


\begin{document}

\maketitle
\newcommand\blfootnote[1]{%
\begingroup
\renewcommand\thefootnote{}\footnote{#1}%
\addtocounter{footnote}{-1}%
\endgroup
}
\begin{abstract}
This paper explores enabling large language models (LLMs) to understand spatial information from multichannel audio, a skill currently lacking in auditory LLMs.
By leveraging LLMs' advanced cognitive and inferential abilities, the aim is to enhance understanding of 3D environments via audio.
We study 3 spatial audio tasks: sound source localization (SSL), far-field speech recognition (FSR), and localisation-informed speech extraction (LSE), achieving notable progress in each task.
For SSL, our approach achieves an MAE of $2.70^{\circ}$ on the Spatial LibriSpeech dataset, substantially surpassing the prior benchmark of about $6.60^{\circ}$. 
Moreover, our model can employ spatial cues to improve FSR accuracy and execute LSE by selectively attending to sounds originating from a specified direction via text prompts, even amidst overlapping speech. 
These findings highlight the potential of adapting LLMs to grasp physical audio concepts, paving the way for LLM-based agents in 3D environments.
\end{abstract}

\section{Introduction}
Empowering large language models \cite{gpt4,llama,vicuna2023} with multimodal perception abilities has emerged as a popular yet challenging research area nowadays. This burgeoning field of research emphasises the integration of LLMs with encoders capable of processing multimodal inputs, including image \cite{flamingo,blip2,instructblip}, 
silent video \cite{lavila} and audio \cite{huck2023emnlp,wu2023decoder,mingqiu2023slm,gong2023whisper,fathullah2023prompting,zhehuai2023salm,yuwenyi2023icassp,tangchangli2023icassp,gong2024listen,salmonn}. 
The development of connection modules and LLM adaptors plays a pivotal role in aligning the encoder output spaces with the LLM input spaces. Such integrated multimodal LLMs are typically trained through cross-modal pre-training and instruction tuning \cite{wei2022instruction,flant5,Ouyang0JAWMZASR22}.\blfootnote{$^*$Equal contribution}\blfootnote{$^\clubsuit$Corresponding author}

Despite the success of multimodal LLMs in managing diverse non-spatial audio and visual tasks, they currently lack the ability to process spatial audio with precise localisation of the sound source in 3D space. In contrast, humans, endowed with binaural hearing, possess the ability to identify sounds, gauge their distance and direction, and selectively focus on sounds originating from a specific direction. This ability allows us to locate a specific person talking in the room and distinguish his/her location from any other speech or sound source. These abilities involve spatial auditory perception and reasoning, which is under-explored in auditory LLMs.


To fill the gap in auditory abilities between humans and LLMs, this paper proposes a spatial audio perception approach in auditory LLMs for 3D spatial speech localisation and recognition. Four-channel spatial audio recorded in the first-order ambisonics (FOA) format is used in our experiments. 
Instead of using speech and audio encoders dedicated to spatial audio processing, we simply use the Whisper \cite{gong2023whisper} speech encoder to encode the semantic content in the omnidirectional signal while providing the intensity vectors (IV) \cite{iv} derived from the B-format audio as spatial information to the LLM. Moreover, a window-level Q-Former \cite{tangchangli2023icassp} is used as a modality aligner to connect the Whisper encoder and the LLM. 


Three spatial speech tasks including 3D sound source localisation (SSL), far-field speech recognition (FSR) and localisation-informed speech extraction (LSE) are studied in this paper. The first two tasks are performed using the Spatial LibriSpeech dataset \cite{spatial_librispeech2023}, which is a spatially augmented synthetic version of LibriSpeech \cite{panayotov2015librispeech} with only one speech source in each sample. The performance of concatenating IV with audio features before and after Q-Former and expanding the vocabulary of the LLM by treating angles as special tokens were compared on the 3D SSL task. A mean angular error (MAE) of $2.70^{\circ}$ can be achieved by concatenating IV with the Whisper encoder output vectors before the Q-Former, whereas the MAE is about $90^{\circ}$ if no spatial information is fed into the LLM. Regarding FSR, our model demonstrated the ability to reduce word error rates (WERs) with additional spatial information. Regarding LSE, Soundspaces 2.0 \cite{chen22soundspaces2} is used to simulate audio samples with two (possibly overlapped) sound sources from different locations based on LibriSpeech \cite{panayotov2015librispeech}, and the dataset is called Dual-Source Spatial (DSS) LibriSpeech. Results show that our model can extract the target speech from the specified direction successfully across a wide range of overlapping ratios.

\section{Related Work}\label{sec:related}

\subsection{On Auditory LLMs}
Several studies have attempted to extend LLMs to support direct speech inputs with a connection module \cite{huck2023emnlp,wu2023decoder,mingqiu2023slm,fathullah2023prompting,zhehuai2023salm,yuwenyi2023icassp}. When LLM-based speech synthesis is also considered, the LLM output space can be augmented with speech tokens as well, such as 
AudioPaLM \cite{rubenstein2023audiopalm}. Unlike speech, audio event inputs are often treated as fixed-sized spectrogram images that can be processed using visual-language LLM methods without explicitly modelling temporal correlations \cite{gong2023whisper,gong2024listen}. Moreover, audio-LLMs have been extended to process speech and audio together 
in an end-to-end fashion for generic hearing abilities \cite{tangchangli2023icassp,salmonn}. 
A contemporary paper to this paper proposes using auditory LLM for question-answering-based SSL for audio events \cite{BAT}. 
However, none of these models can perform SSL based on spatial audio or hear selectively based on the location of the sound source, highlighting the novelty and necessity of this paper.


\subsection{On 3D SSL}

For non-speech audio events, the precise localisation and interpretation of sound sources 
\cite{L3DAS22} are also important. Researchers have innovated with acoustic simulation techniques and developed algorithms that exploit spatial audio's potential 
\cite{simulation_1,simulation_2}.
Specifically, the sound event localisation and detection (SELD) task was introduced in \cite{seld}, which was adopted by the Detection and Classification of Acoustic Scenes and Events (DCASE) community where a range of approaches was designed \cite{dcase_1,dcase_2} 

\subsection{On Spatial FSR}
Spatial information is crucial for many far-field speech processing tasks, such as source separation \cite{sep_tdoa}, speaker diarisation \cite{dia_tdoa_1,dia_tdoa_2} and overlapped speech recognition \cite{he2020spatial}. Meanwhile, efforts have been put into augmenting spatial speech with location annotations via simulation. The work in \cite{libricss} proposed a simulation of overlapped spatial speech datasets for continuous speech separation under a simplified setting with fixed speaker locations. On the other hand, \cite{he2020spatial} proposed a simulation of varying locations of single speakers with noise interference.

\section{Methods}\label{sec:method}

\subsection{Model Structure}
\begin{figure}
    \centering
    \includegraphics[width=0.47\textwidth]{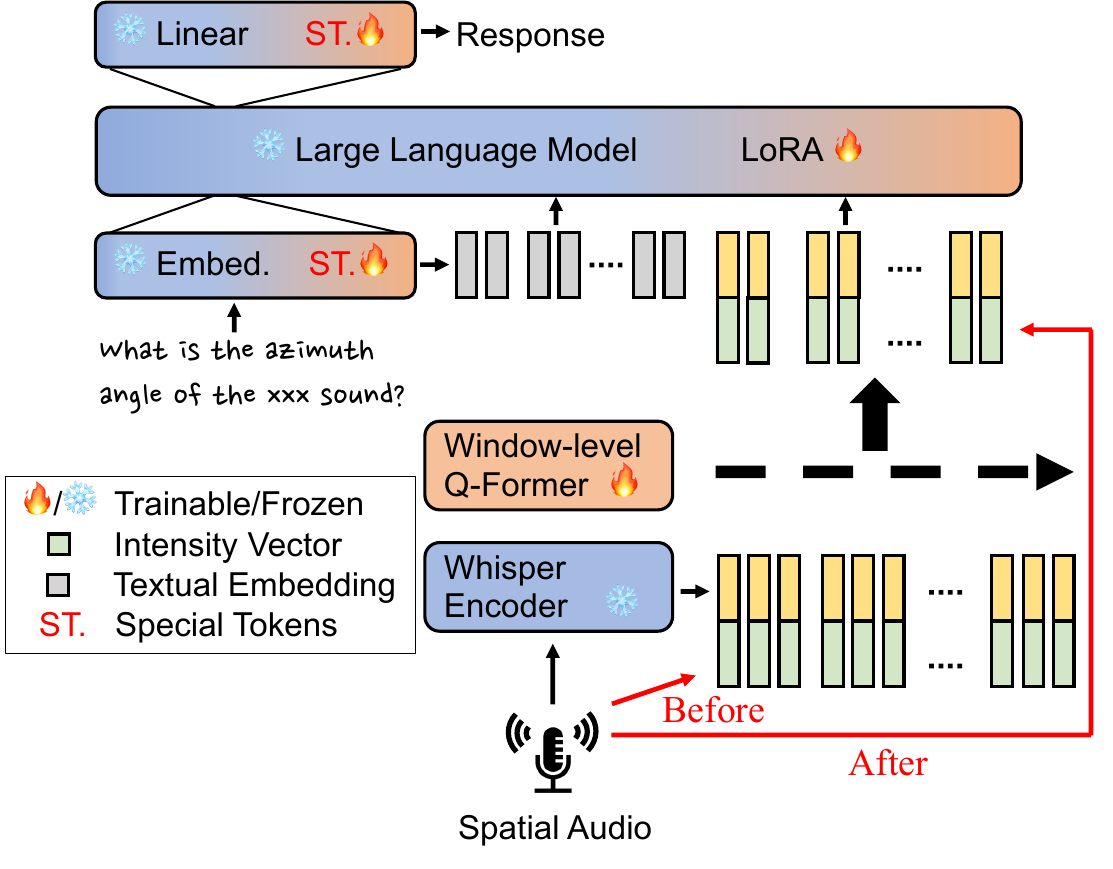}
    \caption{The model structure is shown above. There are two options for introducing spatial information, adding intensity vectors before or after the Q-Former, respectively. Numbers can be also added to the LLM vocabulary as special tokens (ST.), optionally.}
    \label{fig:enter-label}
\end{figure}

Our model architecture is shown in Fig.~\ref{fig:enter-label}, which contains a Whisper \cite{gong2023whisper} encoder, a modality aligner and an LLM with the LoRA module \cite{hu2021lora}. The model accepts spatial audio as input and gives the corresponding textual responses based on textual prompts. Audio recorded in the first-order ambisonics (FOA) format is considered here and the mono-channel signal from the omnidirectional microphone is used as input to the Whisper encoder. To compensate for the lack of spatial information in the omnidirectional signal, intensity vectors (IVs) \cite{iv} calculated from the B-format audio are provided for the LLM. Specifically, the IVs can be introduced after either the Whisper encoder or the modality aligner.

Let $\mathbf{X}$ be the input mono-channel omnidirectional signal and $\mathbf{I}$ be the intensity vectors. With the Whisper encoder $\text{Enc}(\cdot)$, the hidden auditory feature $\mathbf{Z}$ can be obtained as follows:
\begin{equation}
    \mathbf{Z} = \text{Enc}(\mathbf{X}).
\end{equation}
The ``Before'' option introduces the IVs $\mathbf{I}$ before the modality aligner. By controlling the frame rates of $\mathbf{I}$ and $\mathbf{Z}$ to be the same, the two features can be concatenated frame by frame. The obtained feature $\mathbf{Z'}$ will then be fed into the modality aligner and get hidden feature $\mathbf{H}$, as shown in Eqn.~\eqref{equ:option1} and \eqref{equ:align}:
\begin{equation}
\begin{cases}
    \mathbf{Z'} = \text{Concat}(\mathbf{Z}, \mathbf{I}), & \text{``Before''} \\
    \mathbf{Z'} = \mathbf{Z}, & \text{otherwise}
\end{cases},
\label{equ:option1}
\end{equation}
\begin{equation}
    \mathbf{H} = \text{Aligner}(\mathbf{Z'}).
    \label{equ:align}
\end{equation}
The ``After'' option concatenates the IVs to $\mathbf{H}$ after the modality aligner. 
Since the frame rate of $\mathbf{I}$ does not match that of $\mathbf{H}$, the interpolation method is employed here to keep their frame rate the same. Let $\mathbf{I'}$ be the vectors interpolated from $\mathbf{I}$.
After concatenating $\mathbf{I'}$ to $\mathbf{H}$ there is a linear mapping to obtain text-like tokens $\mathbf{H'}$, which are then fed into the LLM, as Eqn.~\eqref{equ:delay} shows.
\begin{equation}
\begin{cases}
    \mathbf{H'} = \text{Linear}(\text{Concat}(\mathbf{H}, \mathbf{I'})), & \text{``After''} \\
    \mathbf{H'} = \text{Linear}(\mathbf{H}) &\text{otherwise}
\end{cases}.
\label{equ:delay}
\end{equation}

Given a text prompt $\mathbf{P}$, the final output textual response $\mathbf{\hat{Y}}$ is generated by the LLM based on $\mathbf{P}$ and $\mathbf{H'}$:
\begin{equation}
    \mathbf{\hat{Y}} = \arg\max\nolimits_\mathbf{Y}P(\mathbf{Y}|\mathbf{P}, \mathbf{H'}).
\end{equation}
The cross-entropy loss is computed based on $P(\mathbf{Y}|\mathbf{P}, \mathbf{H'})$ during training. Both the Whisper encoder and the LLM are frozen. Only the modality aligner and the LoRA structure are updated.

Since the angle is always a number, except tokenising the angles with the tokeniser of the LLM, an optional idea is to expand the vocabulary of the LLM and treat the numbers as special text tokens. Suppose $n$ and $d$ are the original size and text embedding dimension of the vocabulary and $m$ is the number of special tokens to be added.
Let the original vocabulary embedding of the LLM be $\mathbf{V}\in\mathcal{R}^{n\times d}$ and the expanded vocabulary embedding be $\mathbf{V}_\text{new}\in\mathcal{R}^{m\times d}$. We freeze $\mathbf{V}$ and only update $\mathbf{V}_\text{new}$. Due to the change in the vocabulary size, the final linear layer of the LLM has to be expanded. That is, we add a small new linear layer $\mathbf{W}_\text{new}\in\mathcal{R}^{h\times m}$ next to the original linear layer $\mathbf{W}\in\mathcal{R}^{h\times n}$. Similarly, $\mathbf{W}$ is frozen and $\mathbf{W}_\text{new}$ is learnable.


\subsection{Training Strategy}
Question-answering (QA) format data was used for all the tasks. To predict the direction-of-arrivals (DoAs) given by azimuth angle and elevation angle, the text prompt can be \textit{``What is the azimuth angle of the `Male speech' sound?"}, and the model is required to directly output a number. For each audio, both questions about the azimuth angle and the elevation angle were asked to determine the DoA. Similarly, prompts for speech recognition and localisation informed extraction can be \textit{``Please transcribe the speech (on your right) into a written format. "}



\subsection{The DSS LibriSpeech Dataset}
To compensate for the fact that Spatial LibriSpeech only consists of non-overlapped speech \cite{spatial_librispeech2023} and introduce more flexible and diverse sound source locations than LibriCSS \cite{libricss}, we simulated a DSS LibriSpeech dataset. Specifically, the state-of-the-art audio simulator, Soundspaces 2.0 \cite{chen22soundspaces2} was used to provide highly realistic room impulse responses (RIRs). As for the environment meshes, we utilised detailed enough mesh renderings of 90 buildings from Matterport3D \cite{mp3d}. Each building consists of 24.5 rooms across 2.61 floors on average.

For each room, we tried to place the receiver at a navigable coordinate, while for the two sound sources, we placed them within the same room with the receiver. As will be mentioned in Sec. \ref{sec:exp}, sound source locations for two difficulty levels were considered. A simple setting was that the two sound sources were placed on the left or right side of the receiver, noted as \texttt{left/right} dataset. The left side means the DoAs from the receiver is azimuth angle $\phi \in [60^\circ,120^\circ]$ and elevation angle $\theta \in [-30^\circ,30^\circ]$. The right side means the DoAs from the receiver is azimuth angle $\phi \in [-120^\circ,-60^\circ]$ and elevation angle $\theta \in [-30^\circ,30^\circ]$. Another more complicated setting was that the two sound sources were placed randomly, noted as \texttt{random} dataset, which puts higher requirements on the model's spatial angle resolution capability. As for the distances between the sound sources and the receiver, we just randomly placed the sound sources as long as they were in the same room.

For the training sets, we tried to place the receiver at 10 different navigable locations in each room and generated 26704 pairs of RIRs (\texttt{left/right} dataset and \texttt{random} dataset each has 13352 pairs), which were then convolved with randomly sampled utterances from LibriSpeech \cite{panayotov2015librispeech} \texttt{train-clean-100h} subset to finally get a non-overlapped version \texttt{random} dataset by activating the two sources sequentially and another ``fully" overlapped version \texttt{left/right} dataset by activating the two sources simultaneously.\footnote{The generated non-overlapped dataset cannot be treated equally as a concatenated version of Spatial LibriSpeech \cite{spatial_librispeech2023}, because the two utterances are activated under the same reverberation environment.

Here, ``fully" does not mean that the overlapping ratio of the two utterances is 100\% because the durations of them may be different.\label{fn_1}} Each training set consists of about 50k samples. 
For the test sets, we generated several datasets each containing 2k samples with a varying degree of overlapping ratio by sampling utterances from LibriSpeech \texttt{test-clean} subset using the method above.



\section{Experimental Setup}\label{sec:exp_setup}
\subsection{Model Specifications}
We adopt the Whisper-large-v3 \cite{gong2023whisper} encoder as the audio encoder, the window-level Q-Former \cite{tangchangli2023icassp} as the modality aligner, and the Vicuna-7b-v1.5 model \cite{vicuna2023} as the LLM.

For intensity vectors, the hyper-parameters for Short-Time Fourier Transform (STFT) were set to window length=$800$ and hop length=$320$, which generated IVs at the same frame rate 50Hz as the output of the Whisper encoder. Following the best setting in \cite{tangchangli2023icassp}, we utilised Q-Former with a window slide of 0.33s. For LoRA, the rank and the scaling factor were set to 8 and 4.0, respectively. As for expanding the LLM's vocabulary, we added integers from $-180$ to $180$ as special text tokens.


\subsection{Data Specifications}
To compare different model structures without consuming too much resource, experiments in Sec.~\ref{sec:exp_methods} and ~\ref{sec:exp_asr} were performed with randomly selected 51k samples (approximately 30\% of the total data) from the training set of Spatial LibriSpeech \cite{spatial_librispeech2023}, with 1k samples designated for the validation set, and tested on the official Spatial LibriSpeech test set. In Sec.~\ref{sec:exp_asr}, we also collected samples from 
the original LibriSpeech dataset \cite{panayotov2015librispeech} for a close-talk version of the same data used in Spatial LibriSpeech to train a close-talk version system. The proposed DSS LibriSpeech was used in Sec.~\ref{sec:extract}.




\subsection{Task Specifications}
To explore whether LLM can understand spatial speech well, three tasks are focused on in our experiments. Sec.\ref{sec:exp_methods} investigates the use of LLM for 3D SSL while finding the best way to fuse spatial information. Sec.\ref{sec:exp_asr} analysed FSR and LSE was investigated further in Sec.\ref{sec:extract}.

For all the experiments, we set the batch size to 16 and trained the models with 8 A100 GPUs. Models in Sec.\ref{sec:exp_methods} and Sec.\ref{sec:exp_asr} were trained for no more than 30,000 steps and the best checkpoint on the validation set was used for testing. Moreover, models in Table~\ref{tab:non-overlap} were trained for 15,000 steps on the non-overlapped \texttt{random} dataset and by continuing training this model on the ``fully" overlapped \texttt{left/right} dataset for another 15,000 steps, we got the models in Fig.~\ref{fig:ratio}.

\section{Experimental Results}\label{sec:exp}
\subsection{3D Sound Source Localisation (SSL)}
\label{sec:exp_methods}
We first tried introducing spatial information to allow the model to determine the direction of the sound source. Different methods are compared here, including training without any spatial information (w/o IV), using intensity vectors with either the ``Before'' or ``After'' option, and using ``Before'' and tokenising numbers as special tokens.

To evaluate the performance of SSL, \textit{i.e.} decoding the exact orientation of the sound event, the average differences of the azimuth angle, the elevation angle and the angular distance are calculated to get $\Delta_a$, $\Delta_e$ and $\Delta_d$ respectively. Note that $\Delta_d$ is the MAE metric. Since it is the angular distance that most truly reflects the accuracy of the prediction, $\Delta_d$ serves as our primary reference metric. The results are shown in Table~\ref{tab:direction}.

\begin{table}[ht]
    \centering
    \caption{Results of different modelling methods for 3D SSL. ``w/IV'' and ``w/o IV'' refer to with or without using IV, and ``ST." refers to using special tokens to tokenise numbers.}
    \begin{tabular}{lccc}
    \toprule
    Method & $\Delta_a^{\circ}\downarrow$ & $\Delta_e^{\circ}\downarrow$ & $\Delta_d^{\circ} \text{(MAE)}\downarrow$ \\
    \midrule
    w/o IV & 90.19 & 8.26 & 90.16 \\
    w/ IV (``Before'') & \textbf{1.60} & \textbf{1.76} & \textbf{2.70} \\
    w/ IV (``After'') & 3.35 & 2.56 & 4.69 \\
    w/ IV (``Before'') \& ST. & 1.88 & 1.85 & 2.97 \\
    \midrule
    Spatial LibriSpeech \cite{spatial_librispeech2023} & - & - & ~~~6.60 \tablefootnote{In \cite{spatial_librispeech2023}, median angular error is used instead of MAE.} \\
    \bottomrule
    \end{tabular}
    \label{tab:direction}
\end{table}

From Table~\ref{tab:direction}, without any spatial information, the model fails to determine the azimuth angle well, leading to significant deviations in the angular distance. On the other hand, once intensity vectors are introduced, the predictive accuracy of the angles increases significantly, where using ``Before'' leads to about 60\% reduction in MAE compared to the official baseline.

In terms of effectiveness between the different methods, the prediction of direction is more accurate using ``Before'' compared to ``After''. This suggests that it is difficult to align the spatial information, such as IVs, with the input textual space of the LLM, using only a single linear transformation like in ``After''. It is possible that a more powerful transformation can align the IVs well with the LLM input space. 
In addition, adding special tokens for numbers to LLM's input embedding slightly impairs the performance. This implies that LLM itself has a good understanding of numbers and doesn't need a separate expanded vocabulary to learn numbers. 
Moreover, LLM may not understand the relative relationships between the newly added number tokens, which is different from the original tokenisation was trained on a large amount of textual corpus, which might explain the drop in performance.

Based on the results, it is proved to be a decent setup using intensity vectors with ``Before'' without adding special number tokens. Therefore, we use this configuration throughout the rest of the experiments, denoted by ``w/ IV".

\subsection{Far-field Speech Recognition (FSR)}
\label{sec:exp_asr}
Only being able to recognise the direction of the sound is not sufficient to achieve full spatial audio understanding of LLMs. It is more necessary to understand the audio content in conjunction with the spatial information. We investigated speech recognition for spatial audio here, and consider three settings: using non-Spatial LibriSpeech data and using Spatial LibriSpeech with and without adding intensity vectors. The results are shown in Table~\ref{tab:asr}.

\begin{table}[ht]
    \centering
    \caption{Results of speech recognition under different settings. ``w/IV'' and ``w/o IV'' refer to with or without using IV.}
    \begin{tabular}{lcc}
    \toprule
    Method & Data & WER\%$\downarrow$\\
    \midrule
    w/o IV & LibriSpeech & 8.6 \\ 
    w/o IV & Spatial LibriSpeech & 9.0 \\
    w/ IV & Spatial LibriSpeech & \textbf{7.6} \\
    \bottomrule
    \end{tabular}
    \label{tab:asr}
\end{table}
Training on spatial audio with intensity vectors turns out to be the best according to the results, indicating that LLM has learnt to use spatial information to assist FSR.

\subsection{Localisation-Informed Speech Extraction (LSE)}\label{sec:extract}
\label{sec:extract}
Three metrics are leveraged to analyse the performance of the localisation-informed speech extraction task. First, we found that audio-LLM cannot always transcribe the speech from the required direction but recognizes the speech from the other direction or outputs a concatenated/mixed version of the two utterances. Thus we propose a GPT-assisted approach \footnote{We inform GPT-3.5-turbo the output of the model, the transcription of the target speech and the transcription of the distracting speech and design a prompt to request it to judge whether the model attempts to transcribe the target speech. LLM-assisted evaluation methods are widely used in NLP and ML. The correctness is based on randomly sampled human checks.} to judge whether the audio-LLM extracts the target speech successfully and reports a success rate (SR). Then we calculate WERs for both the successful samples only and all the samples, noted as sWER and WER respectively, to evaluate the accuracy of the model in recognizing target speech.

We first analysed whether spatial information is crucial for LSE using a non-overlapped version dataset. As shown in Table~\ref{tab:non-overlap}, if any spatial information is not provided for the LLM, the model achieves SR\% around 50 on both \texttt{left/right} and \texttt{random} datasets, which means the model cannot extract the speech from the target direction as required. On the contrary, models with IV as input can extract the target speech successfully at a SR\% over 80, which demonstrates the importance of spatial information in this task.

Moreover, Table~\ref{tab:non-overlap} shows that the model's performance on \texttt{random} dataset is worse than that on \texttt{left/right} dataset, which reveals that the closer the angular distance between the two sound sources is, the harder it is for the model to distinguish the two utterances from each other.

\begin{table}[ht]
    \centering
    \caption{Results on the non-overlapped version of DSS LibriSpeech. ``w/ IV'' and ``w/o IV'' refer to with or without using IV. Performance on \texttt{left/right} dataset and \texttt{random} dataset are compared.}
    \begin{tabular}{lcccc}
    \toprule
    Method & Data & SR\%$\uparrow$ & sWER\%$\downarrow$ & WER\%$\downarrow$ \\
    \midrule
    \multirow{2}{*}{w/o IV} & left/right & 50.1 & 4.4 & 67.2 \\
    ~ & random & 50.0 & 4.0 & 67.5 \\
    \midrule
    \multirow{2}{*}{w/ IV} & left/right & 96.3 & 5.1 & 9.0 \\
    ~ & random & 80.2 & 5.4 & 29.8 \\
    \bottomrule
    \end{tabular}
    \label{tab:non-overlap}
\end{table}

\begin{figure}[ht]
    \begin{minipage}[b]{.48\linewidth}
    \centering
        \centerline{\includegraphics[width=4.0cm]{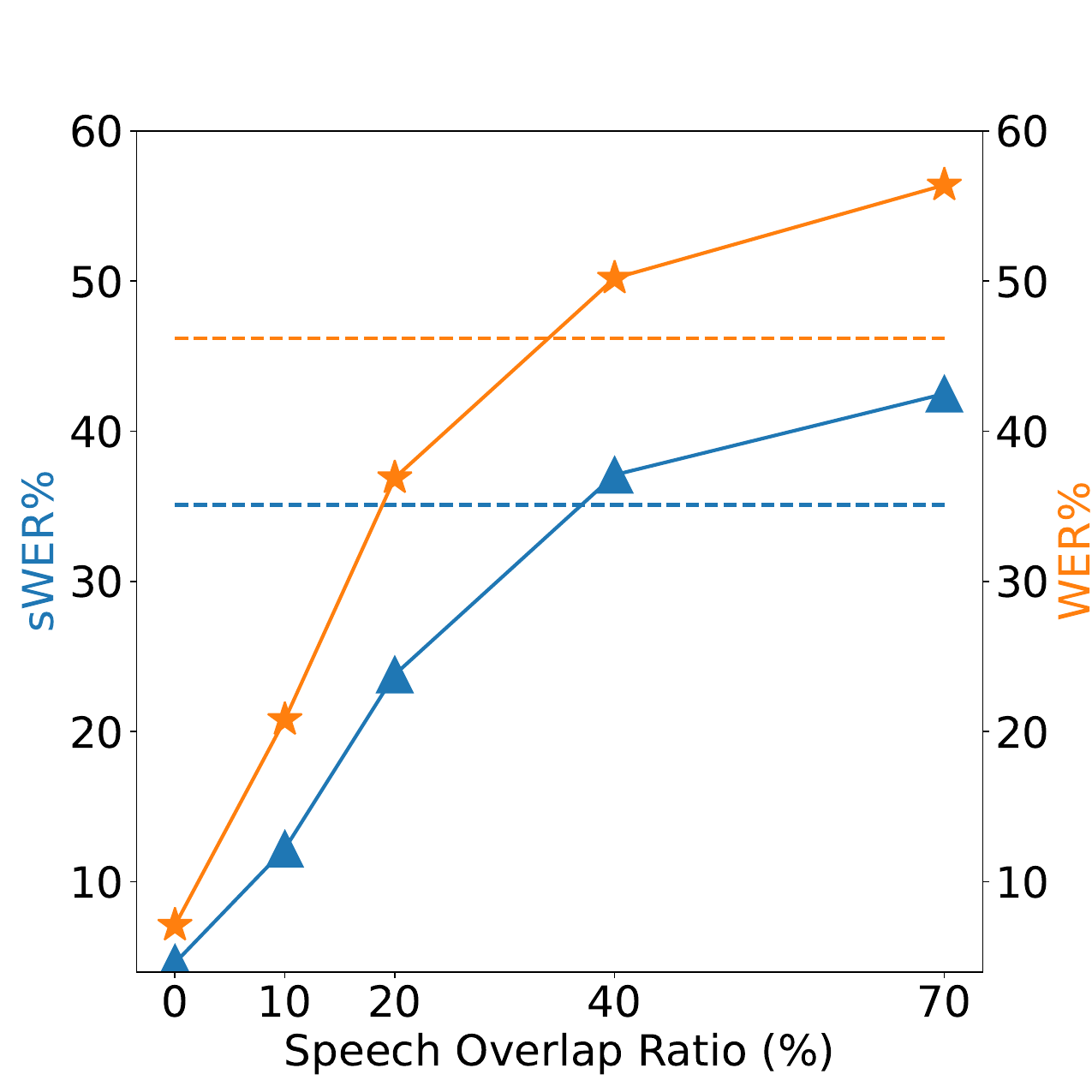}}
    \end{minipage}
    \hfill
    \begin{minipage}[b]{.48\linewidth}
    \centering
        \centerline{\includegraphics[width=4.0cm]{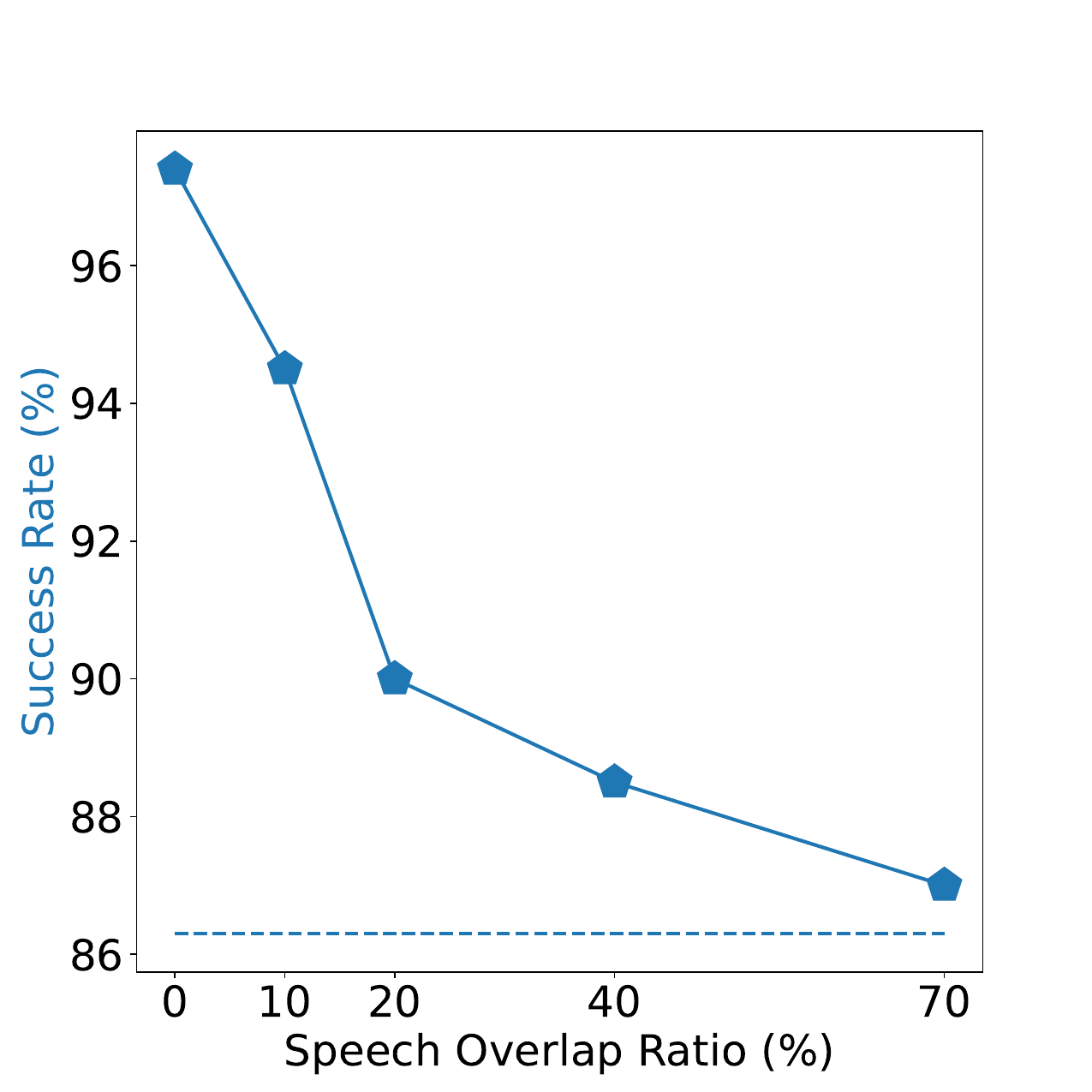}}
    \end{minipage}
    \caption{Results on the \texttt{left/right} dataset with overlapping ratio from 0\% to 70\%. The dotted lines are the performance on the ``fully" overlapped test set generated by activating the two sources simultaneously. (as described in Footnote~\ref{fn_1})}
    \label{fig:ratio}
\end{figure}

Next, we used the \texttt{left/right} dataset to investigate the impact of the overlapping ratio of the two utterances. As shown in Fig.~\ref{fig:ratio}, with the increasing of the overlapping ratio, the model's performance gradually degrades. It shows that the model still struggles to extract the specified speech under circumstances where utterances are highly overlapped with each other. This performance degradation is attributed to both the reduction in success rate and the deterioration in WERs.

\section{Conclusion}\label{sec:conclusion}
This paper delves into the vital exploration of enabling LLMs to understand spatial audio. 
Experiments show that introducing spatial information, such as intensity vectors, before the modality aligner is suitable for the auditory LLMs to perceive spatial audio well, and expanding the LLMs' vocabulary with direction-related special tokens can impair the performance. 
Advanced results are obtained in SSL with an MAE of $2.70^{\circ}$, which is a nearly 60\% improvement over the official Spatial LibriSpeech baseline using only 30\% of all training samples in the dataset. 
Meanwhile, the pivotal role of spatial information in enhancing LLMs' comprehension of spatial audio in FSR is also underscored. 
In addition, an overlapped spatial speech dataset is simulated, and our model can selectively attend to audio in different directions based on different text prompts, showing the potential of future artificial intelligence to perceive and understand real-world sounds like humans.

\newpage
\bibliographystyle{IEEEtran}
\bibliography{mybib}

\end{document}